\documentclass[12pt]{article}

\usepackage{scicite}
\usepackage{graphicx, amsmath, amssymb, caption}

\usepackage{times}

\newenvironment{sciabstract}{%
\begin{quote} \bf}
{\end{quote}}

\newcounter{lastnote}
\newenvironment{scilastnote}{%
\setcounter{lastnote}{\value{enumiv}}%
\addtocounter{lastnote}{+1}%
\begin{list}%
{\setlength{\leftmargin}{.22in}}
{\setlength{\labelsep}{.5em}}}
{\end{list}}

\title{KOI-3278: A Self-Lensing Binary Star System}

\author
{Ethan Kruse$^{1\ast}$ and Eric Agol$^{1}$ \\
\\
\normalsize{$^{1}$Department of Astronomy, University of Washington,}\\
\normalsize{Box 351580, Seattle, WA 98195, USA.}\\
\\
\normalsize{$^\ast$Corresponding author. E-mail:  eakruse@uw.edu}
}

\date{}

\begin{document} 


\baselineskip12pt


\maketitle 

\begin{sciabstract}
Over 40\% of Sun-like stars are bound in binary or multistar systems. 
Stellar remnants in edge-on binary systems can gravitationally 
magnify their companions, as predicted 40 years ago. By using data 
from the Kepler spacecraft, we report the detection of such a       
``self-lensing'' system, in which a 5-hour pulse of 0.1\% amplitude   
occurs every orbital period.  The white dwarf stellar remnant and  
its Sun-like companion orbit one another every 88.18 days,     
a long period for a white dwarf-eclipsing binary. By modeling the     
pulse as gravitational magnification (microlensing) along with  
Kepler's laws and stellar models, we constrain the mass of the    
white dwarf to be $\sim$63\% of the mass of our Sun. Further     
study of this system, and any others discovered like it, will help 
to constrain the physics of white dwarfs and binary star evolution.

\end{sciabstract}

\section*{}

Einstein's general theory of relativity predicts that gravity can bend
light and, consequently, that massive objects can distort and magnify
more distant sources \cite{Einstein1936}. This lensing effect provided
one of the first confirmations of general relativity during a solar
eclipse \cite{Dyson1920}. Gravitational lensing has since become a
widely used tool in astronomy to study galactic dark matter,
exoplanets, clusters, quasars, cosmology, and more
\cite{Wambsganss1998,Meylan2006}. One application has yet to be
realized: in 1973, Andr\'e Maeder predicted that binary star systems
in which one star is a degenerate, compact object -- a white dwarf,
neutron star, or black hole -- could cause repeated magnification of
its companion star (instead of the standard eclipses) if the orbit
happened to be viewed edge-on \cite{Maeder1973}. The magnification of
these self-lensing binary systems is small, typically a part in one
thousand or less if the companion is Sun-like, and so it was not until
high-precision stellar photometry was made possible with the Corot and
Kepler spacecrafts that this could be detected
\cite{Agol2002,Sahu2003}. Stellar evolution models predict that about
a dozen self-lensing binaries could be found by the Kepler spacecraft
\cite{Farmer2003}, but none have been discovered to date. A
self-lensing binary system allows the measurement of the mass of the
compact object, which is not true for most galaxy-scale microlensing
events in which there is a degeneracy between the velocity, distance,
and mass of the lensing object \cite{Paczynski1996}. Microlensing does
affect several known white dwarfs in binaries in which the depth of
eclipse is made slightly shallower
\cite{Marsh2001,Steinfadt2010,Rowe2010,Muirhead2013,Kaplan2014} but
does not result in brightening because occultation dominates over
magnification at the short orbital periods of those systems.

\begin{figure*}
\center
\includegraphics[width=7in]{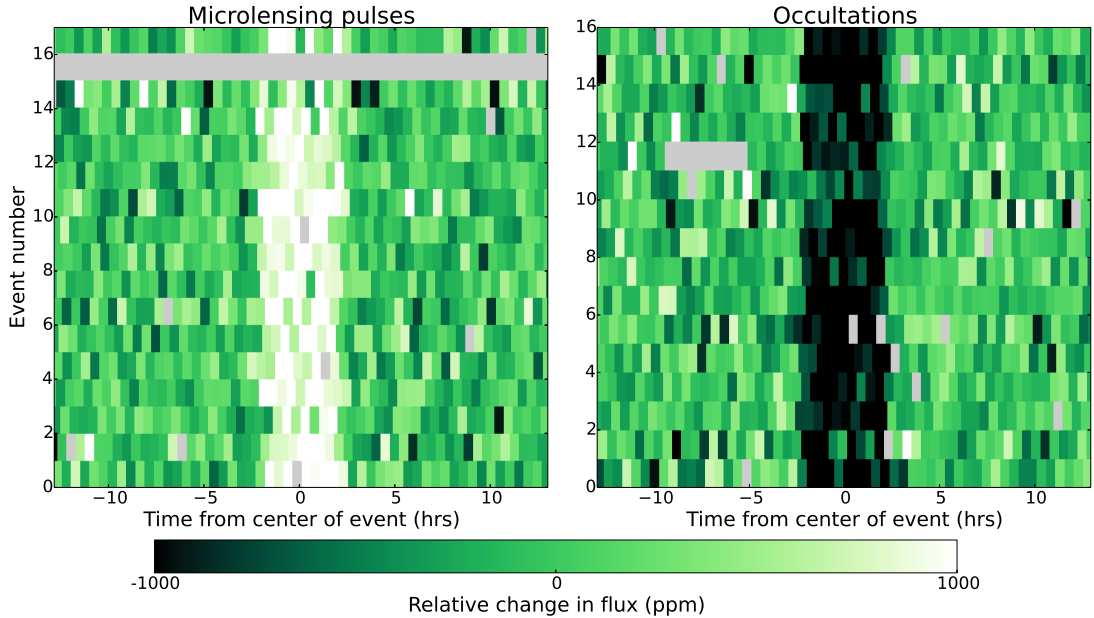}
\caption*{
 {\bf Fig. 1. Detrended flux versus time for all 16 microlensing pulses and 16 occultations in KOI-3278}. Each row depicts the relative fluxes in 29.3-min Kepler cadences around an event. The rows are
separated by the orbital period, $P = 88.18$ days. White represents brighter flux and black dimmer, whereas
gray represents missing data or outliers that have been removed. ppm, parts per million.
  }
\end{figure*}

Here, we report that Kepler Object of Interest 3278 (KOI-3278)
\cite{Burke2014,Tenenbaum2014}, a term intended for planetary
candidates, is instead a self-lensing binary composed of a white dwarf
star orbiting a Sun-like star. The candidate planetary transit signal
is actually the repeated occultation of the white dwarf as it passes
behind its stellar companion.  A search for other planets in this
system with the Quasiperiodic Automated Transit Search algorithm
\cite{Carter2013} turned up a series of symmetric pulses, brightenings
rather than dimmings, with a near-identical period and duration as the
transit candidate but occurring almost half an orbital period
later. All these properties can be explained by magnification of the
Sun-like star as the white dwarf passes in front; 16 microlensing
pulses were found, in addition to 16 occultations. The pulses and
occultations are periodic and uniform in magnitude and duration
(Fig. 1), which is consistent with a nearly circular, Keplerian
orbit. Because there is no other phenomenon (that we know of) that
can cause such a brief, symmetric, periodic brightening, we
constructed a model for KOI-3278 composed of an eclipsing white dwarf
and G dwarf (Sun-like star) binary \cite{SOM}.

Even without a full model, an estimate of the
mass of the white dwarf, $M_2$, can be made directly
from the light curve. The ratio of the fluence of
the microlensing pulse, $F_\text{pulse}$, to the stellar fluence
over an orbital period, $F_\text{tot}$, is given \cite{Agol2003}
by $F_\text{pulse}/F_\text{tot} = 5.4 \times 10^{-6} \sqrt{\left(1-b^2\right)} (M_2/M_\odot) (R_\odot/R_1)$, where
$R_1$ is the radius of the G dwarf and $b$ is the
impact parameter \cite{foot1}. Because the duration of
the pulse is 5 hours, the period is 88.18 days,
and the magnification is $10^{-3}$, we calculated
$F_\text{pulse}/ F_\text{tot}[1-(b/0.7)^2]^{-1/2} \approx 3.3 \times 
10^{-6}$ and $M_2 \approx 0.6 M_\odot$, which is a typical mass for a white
dwarf star \cite{Kepler2007}.

\begin{figure*}
\center
\includegraphics[width=7.in]{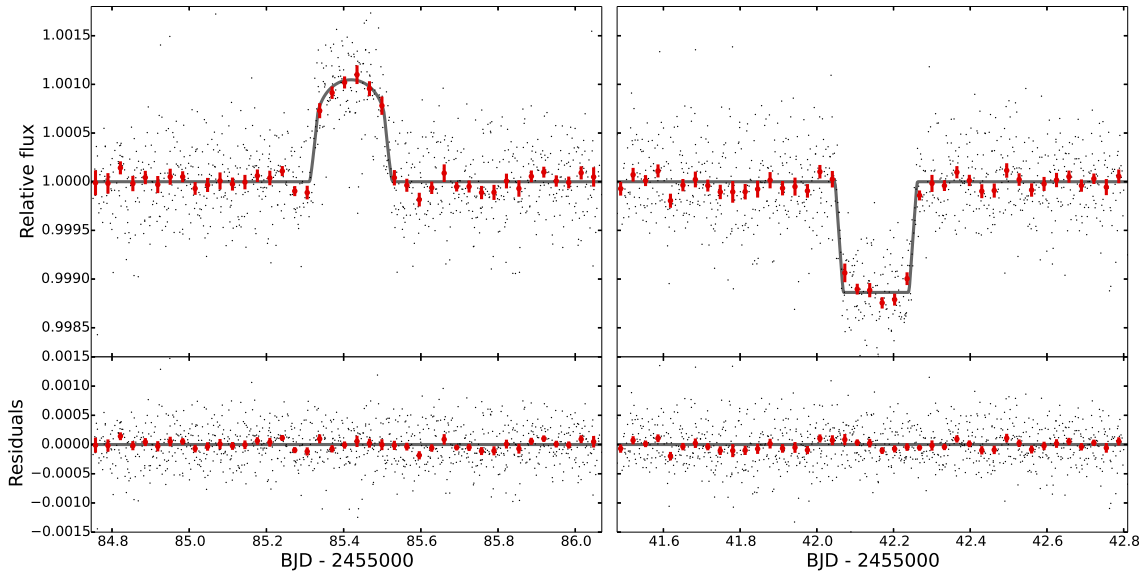}
 \caption*{
{\bf Fig. 2. Model fit to the data.} Detrended and folded Kepler photometry of
KOI-3278 presented as black points (all pulses and occultations have been
aligned), overplotted with the best-fit model (gray line) for the microlensing
pulse ({\bf left}) and occultation ({\bf right}). Red error bars show the mean of the
folded data over a 45-min time scale. Bottom graphs show the residuals of the
data with the best-fit model subtracted. BJD, barycentric Julian date.
}
\end{figure*}

To jointly constrain the parameters of both
stars, we fitted a full model simultaneously to the
Kepler time-series photometry and the multiband
photometry collected from other surveys \cite{SOM}.
We modeled the light curve by using a Keplerian
orbit with the gravitational lensing approximated
as an inverted transit light curve, which is appropriate
when the Einstein radius is small \cite{Agol2003}.
We compared the Padova stellar evolution models
\cite{Bressan2012} to the multiband photometry to constrain
the properties of the G dwarf while accounting
for extinction, $A_\text{V}$. Last, we used cooling models to
constrain the age of the white dwarf \cite{Bergeron2011}.

Our model provides an accurate description
of the data (Fig. 2) with a reduced $\chi^2$ value of
unity. From this model, we calculated the stellar
parameters and the binary system's orbital properties (Table 1),
with uncertainties derived from a Markov-chain
Monte Carlo analysis (18). The model produced
a white dwarf mass of $M_2 =0.63 M_\odot \pm0.05 M_\odot$,
with a G dwarf companion of $M_1 = 1.04^{+0.03}_{-0.06} M_\odot$,
 $R_1 =0.96^{+0.03}_{-0.05} R_\odot$,
and effective temperature $T_{\text{eff},1} = 5568\pm 39$ K: a star very similar to our
Sun. Because the white dwarf, with its small size,
is much fainter than the G dwarf, we cannot yet
measure its temperature directly. However, given
the measured mass from gravitational lensing, we
inferred its size to be $R_2 = 0.0117 R_\odot \pm0.0006 R_\odot$
by using a mass-radius relation appropriate for
carbon-oxygen white dwarfs. With a radius for
the white dwarf, the measured occultation depth
when it passes behind the G dwarf can be used to
constrain the temperature of the white dwarf,
which we found to be $T_{\text{eff},2} = 10,000\pm 750$ K;
this temperature would give the white dwarf the
bluish hue of an A star. The Einstein radius, $R_\text{E}$, is
about twice the inferred size of the white dwarf,
which allows lensing to dominate over occultation
when the white dwarf passes in front. Gravitational
lensing causes a distorted and magnified
image of the G dwarf outside the Einstein ring in
addition to a second inverted and reflected image 
of the G dwarf within the Einstein ring (Fig. 3);
the inner image is partially occulted by the white
dwarf's disk, reducing the observed magnification
slightly.

\begin{table}
\begin{center}
\caption{ \label{tab:mcmc_results} 
{\bf Parameters of the KOI-3278 binary
star system.} More information can be found in
the supplementary text. The median and 68.3\%
bounds are given for each parameter. 
$g_1$, surface gravity in cm/s$^2$. $L_{WD}$, luminosity of the white dwarf. $e$, eccentricity. $\omega$, argument of periastron. $a$, semi-major axis. $i$, inclination. $F_2/F_1$, flux ratio between the white dwarf and G dwarf in the Kepler band. $D$, distance. $\sigma_{sys}$, systematic errors in the multiband photometry.}
\begin{tabular}{lc}
\hline \hline
\multicolumn{2}{c}{\it G dwarf:}\\
\hline
$M_1 (M_\odot)$ & $1.042^{+0.028}_{-0.058}$ \\
$R_1 (R_\odot)$ & $0.964^{+0.034}_{-0.054}$ \\
$[Fe/H]_1$ & $0.39^{+0.22}_{-0.22}$ \\
$t_1$ (Gyr) & $1.62^{+0.93}_{-0.55}$ \\ 
$T_{eff,1}$ (K) & $5568^{+40}_{-38}$ \\
log($g_1$) & $4.485^{+0.026}_{-0.020}$ \\
\hline \hline
\multicolumn{2}{c}{\it White dwarf:}\\
\hline

$M_{2,init} (M_\odot)$ & $2.40^{+0.70}_{-0.53}$ \\
$M_2 (M_\odot)$ & $0.634^{+0.047}_{-0.055}$ \\
$T_{eff,2}$ (K) & $9960^{+700}_{-760}$ \\
$R_2 (R_\odot)$ & $0.01166^{+0.00069}_{-0.00056}$ \\
$R_E (R_\odot)$ & $0.02305^{+0.00093}_{-0.00107}$ \\
$t_2$ (Gyr) & $0.96^{+0.90}_{-0.53}$ \\
$t_{cool}$ (Gyr) & $0.663^{+0.065}_{-0.057}$ \\
$L_{WD} (L_\odot)$ & $0.00120^{+0.00024}_{-0.00023}$ \\

\hline \hline
\multicolumn{2}{c}{\it Binary system:} \\
\hline

$P$ (d) & $88.18052^{+0.00025}_{-0.00027}$ \\
$t_0$ (-2,455,000 BJD) & $85.4190^{+0.0023}_{-0.0023}$ \\
$e\cos\omega$ & $0.014713^{+0.000047}_{-0.000061}$ \\
$e\sin\omega$ & $0.000^{+0.049}_{-0.054}$ \\
$a$ (AU) & $0.4605^{+0.0064}_{-0.0103}$ \\
$a/R_1$ & $102.8^{+3.7}_{-2.4}$ \\
$b_0$ & $0.706^{+0.020}_{-0.025}$ \\
$i$ (deg) & $89.607^{+0.026}_{-0.020}$ \\
$F_2/F_1$ & $0.001127^{+0.000039}_{-0.000039}$ \\
$D$ (pc) & $808^{+36}_{-49}$ \\
$\sigma_{sys}$ & $0.0246^{+0.0127}_{-0.0078}$ \\
\hline
$K_1$ (km/s) & $21.53^{+0.96}_{-0.98}$ \\
$\pi$ (milli-arc sec) & $1.237^{+0.079}_{-0.053}$ \\
$\alpha_1$ (milli-arc sec) & $0.2169^{+0.0076}_{-0.0072}$ \\
$A_V$ (mags) & $0.206^{+0.017}_{-0.016}$ \\

\hline \hline
\end{tabular}
\end{center}
\end{table}

Our model does not include the effect of star
spots, but the Kepler G dwarf light curve displays
their characteristic quasi-periodic fluctuations with
a root mean square of 0.76\%. We estimated that
the spots would affect the derived stellar properties
by less than a percent, smaller than the statistical
errors in our model. Spot analysis revealed
a G dwarf rotational period of $P_\text{rot}=12.5\pm 0.1$
days. This short rotational period is expected for
a G dwarf of only $0.89\pm0.14$ Gy \cite{SOM}. The white
dwarf cooling time is $t_\text{cool} = 0.66\pm0.06$ Gy, which
when added to the main sequence lifetime, $t_2$, of its progenitor with mass $M_{2,\text{init}}$ gives a
total age of the binary system of $t_1=1.6^{+0.9}_{-0.6}$ Gy; this age is marginally
inconsistent (1.4$\sigma$) with the spin-down age
of the G dwarf.

\begin{figure*}
\center
\includegraphics[width=6.5in]{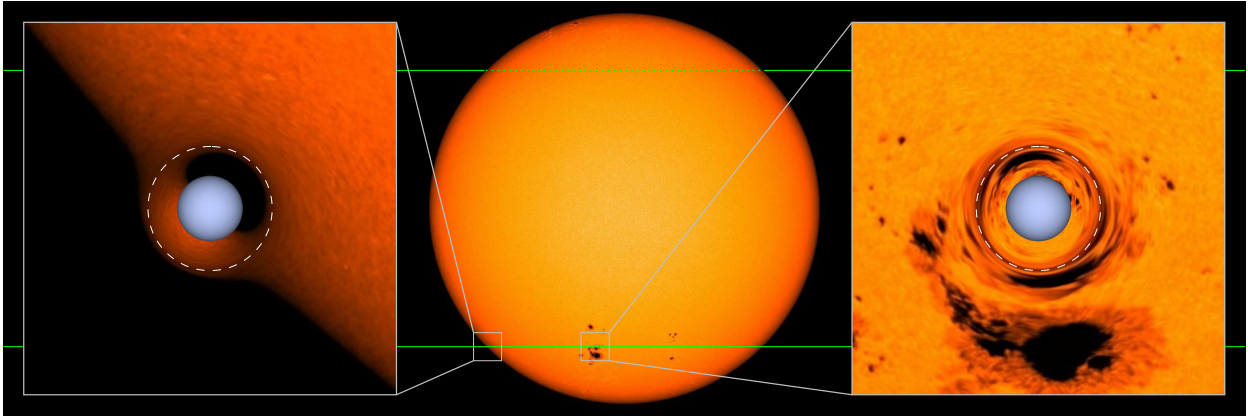}
\caption*{
{\bf Fig. 3. Illustration of lensing magnification. (Center)} The false-color disk of a G dwarf (using an
actual image of the Sun from NASA/SDO HMI), in which the green line shows the trajectory of the white
dwarf, with the dotted portion indicating where it passes behind the G dwarf. ({\bf Left} and {\bf right}) Close-ups
of areas boxed in center show the lensed image of the G dwarf at two different times during the
microlensing pulse; the white dwarf is the blue sphere. The white dashed line shows the Einstein ring of
the white dwarf. The model that we fit to the data does not contain spots; however, the spots and
granulation make the lensing distortion more apparent.
}
\end{figure*}

However, the G dwarf may have been spun
up because of accretion of gas from the white
dwarf progenitor. Because the white dwarf progenitor
was previously a red giant, it should
have enveloped the G dwarf during a common envelope
phase \cite{Ivanova2013}. The initial orbital period
of the binary was likely several years long, and
the period was likely shortened because of drag
during the common-envelope phase. During this
phase, the G dwarf accreted some gas from the
red giant, increasing its mass by $10^{-3}$ to $10^{-2} M_\odot$
and spinning the G dwarf up from the angular
momentum contained in the accreted gas; this
spin-up would have reset the age-spin relation,
which could explain the slight age discrepancy.

KOI-3278 is the longest period eclipsing post-common-envelope
 binary found to date (fig. S7),
and it is also one of the only examples of an
eclipsing Sirius-like system -- a binary composed
of a non-interacting white dwarf and a Sun-like
(or hotter) main-sequence star \cite{Rebassa2012,Zorotovic2013,Holberg2013}. As such,
it will help to provide constraints on the physics
of formation and evolution of short and intermediate
period evolved binary stars, thereby improving
our knowledge of the formation of accreting
binaries and sources of gravitational waves. We
expect that a few more self-lensing binaries will
be found in the Kepler data at shorter orbital periods
than KOI-3278. The magnification decreases
down to periods of $\approx$16 days, making them more
difficult to find; at even shorter periods, occultation
by the white dwarfÕs disk wins out over the
lensing, causing a shallower eclipse as in KOI-256
\cite{Muirhead2013}. Systems like KOI-3278 should not be a substantial
source of false-positives for exoplanets;
only one was predicted to be found in the Kepler
data with its magnification of $\approx$0.1\%\cite{Farmer2003}.

Follow-up observations should better constrain
the parameters of the white dwarf star in KOI-3278,
 allowing a test of the mass-radius relation
for white dwarfs \cite{Provencal1998,Parsons2012}. Once the Kepler field
rises (it had set before we detected the microlensing signal),
  radial velocity observations should show
a semi-amplitude of $K_1 = 21.5$ km/s and a line-broadening
of 4 km/s. High-resolution spectroscopy will also
better constrain the atmospheric properties of the
G dwarf; in particular, spectral abundance anomalies
caused by accretion of nuclear-processed material
from the white dwarf progenitor should
be sought. Measurements of the occultation of the
white dwarf in the ultraviolet (with the Hubble
Space Telescope) should appear much deeper, as
much as 60\% in depth as opposed to the 0.1\%
occultation depth in the Kepler band, and will
yield constraints on the radius and temperature of
the white dwarf. High angular resolution imaging
would allow for better constraints to be placed on
the presence of a third star in the system\cite{SOM}. Last,
parallax measurements, $\pi$, with the Gaia spacecraft
\cite{Perryman2001}  will improve the precision of the properties of
the G dwarf; Gaia can also detect the reflex motion, $\alpha_1$,
of the G dwarf as it orbits the center of mass with
the white dwarf. This provides another means to
detect systems like KOI-3278 with inclinations that
do not show microlensing or occultation; there are
likely 100 of these among the Kepler target stars
alone, given the ~1\% geometric lensing probability
of KOI-3278.

\bibliographystyle{Science}

\bibliography{scibib}

\begin{thebibliography}{100}
\bibitem{foot1} The impact parameter is the projected sky separation of
the white dwarf and G dwarf at midpulse, in units of the
radius of the G dwarf; we find a best-fit value of $b = 0.7$
based on the full model fit. This formula assumes a
circular orbit, neglects limb-darkening, and neglects
obscuration by the white dwarf; whereas these effects
(although minor) are accounted for in our full model.
\end{thebibliography}

\bibliographystyle{Science}

\begin{scilastnote}
\item {\bf Acknowledgments:} E.K. was funded by an NSF Graduate
Student Research Fellowship. E.A. acknowledges funding by
NSF Career grant AST 0645416; NASA Astrobiology InstituteÕs
Virtual Planetary Laboratory, supported by NASA under
cooperative agreement NNH05ZDA001C; and NASA Origins of
Solar Systems grant 12-OSS12-0011. Solar image courtesy of
NASA/Solar Dynamics Observatory (SDO) and the Helioseismic
and Magnetic Imager (HMI) science teams. The Kepler data
presented in this paper were obtained from the Mikulski
Archive for Space Telescopes (MAST). The code used for
analysis is provided in a repository at github.com/ethankruse/koi3278. 
The authors welcome requests for additional
information regarding the material presented in this paper.

{\bf Supplementary Materials}

www.sciencemag.org/content/344/6181/[PAGE]/suppl/DC1

Materials and Methods

Supplementary Text

Figs. S1 to S7

Table S1

References (31--84)

10 February 2014; accepted 25 March 2014

10.1126/science.1251999

\end{scilastnote}

\clearpage

\setcounter{figure}{0}
\setcounter{table}{0}
\renewcommand{\thefigure}{S\arabic{figure}}
\renewcommand{\thetable}{S\arabic{table}}

\paragraph*{\Huge Supplementary Materials} 

\paragraph{Here we provide additional description of our methods in modeling 
and constraining the properties of KOI-3278.  }

\subsection{Terminology}

We use the term ``self-lensing binary'' to refer to a binary star
system that is edge-on and in which one star causes a brightening of 
its companion -- due to gravitational 
magnification, or ``microlensing'' -- as it passes in front of the companion's disk \cite{Gould1995,Sahu2003,Rahvar2011,Mao2012}.
Since a self-lensing binary has not been detected to date, we need to 
define some terminology for periodic microlensing in a binary star
system.   In particular, the brightening that occurs in KOI-3278 when 
the white dwarf magnifies the G dwarf is neither an eclipse nor a transit, 
which are associated with a decrease in the brightness of the system.  Nor
can this be described as a ``microlensing event'' since it repeats;
it is not a single event.  Maeder \cite{Maeder1973} used the term ``gravitational 
flash'' to describe repeated microlensing in a binary;  however, this
term could also connote gravitational waves or explosive events.  Others have 
used the term ``anti-transit''\cite{DiStefano2011}, but this has also
been used to refer to a secondary eclipse that happens opposite in
the orbit to the transit\cite{Winn2011}.  Instead of these terms, we refer to the 
series of brightenings that occur as the white dwarf magnifies the G dwarf 
as a ``microlensing pulse train'', and to a single event as a ``pulse.''

We refer to the G dwarf as the primary star and the white
dwarf as the secondary, and we label their physical properties with
$1$ and $2$; thus the masses and radii are $M_1$ and $R_1$ for
the G dwarf and $M_2$ and $R_2$ for the white dwarf.
We refer to the secondary eclipse, when the white dwarf passes behind
the G dwarf, as the occultation.

\section{Photometric time series model} \label{sec:photometry}

\subsection{Kepler photometry} \label{sec:photometry:kepler}

We used the simple-aperture-photometry flux ({\sc SAP\_FLUX}) from the Kepler
pipeline for all available quarters (Q1-Q17).  The times are the mid-point of each
cadence, converted to barycentric julian date ({\sc BJD}).  We rejected
points that were flagged with cosmic ray contamination or single-point outliers (SAP\_QUALITY flags 128 and 2048).

\begin{figure*}
\center
\includegraphics[width=7.in]{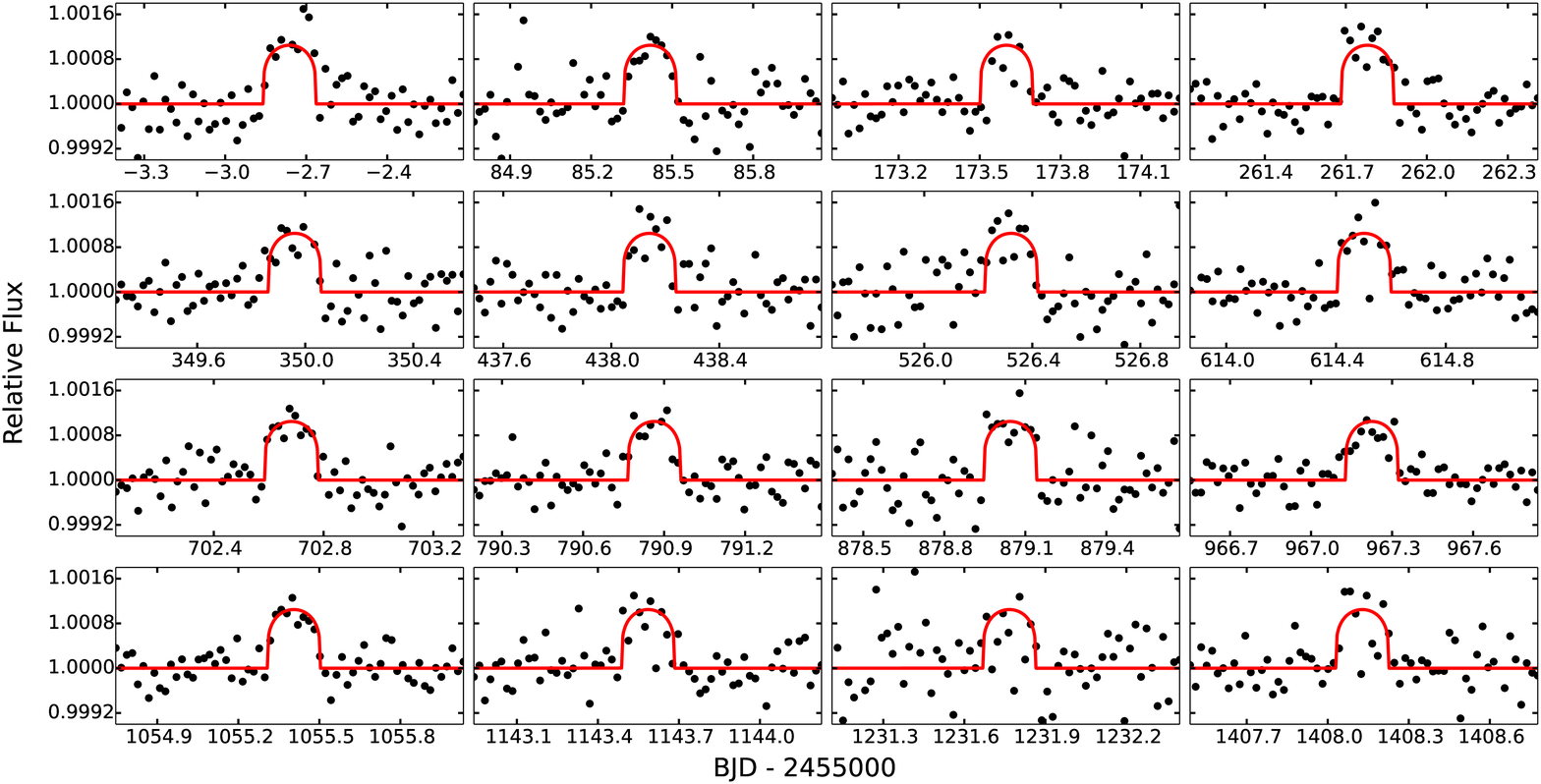}
\includegraphics[width=7.in]{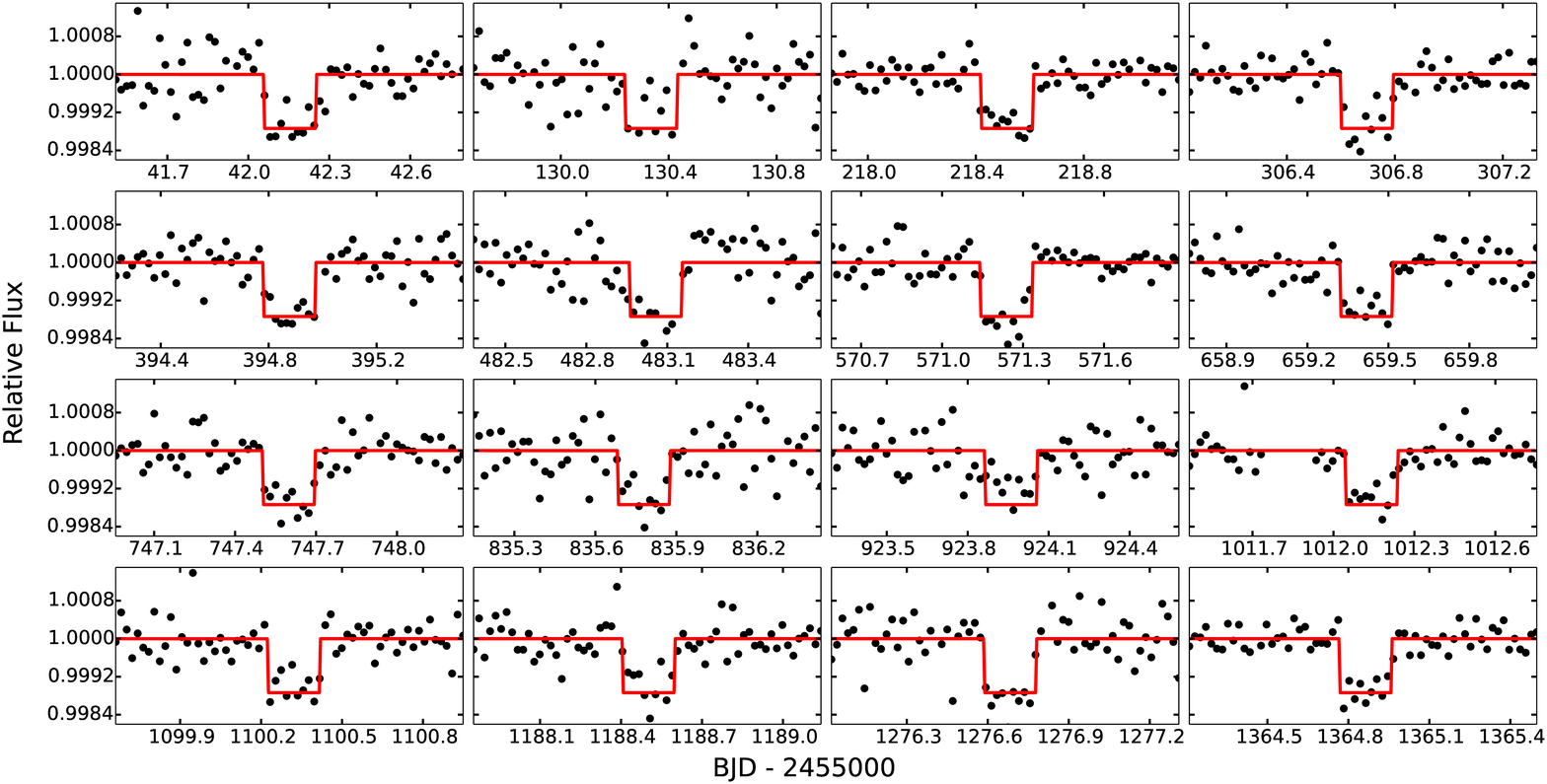}
 \caption{Detrended light curves encompassing the individual
microlensing pulses (top) and occultations (bottom). The red lines show our best-fit model. 
\label{fig:lightcurve_raw}}

\end{figure*}

A plot of the pulses and occultations is shown in Figure \ref{fig:lightcurve_raw}.

\subsection{Light curve model}
\label{lcmodel}

\begin{figure*}
\center
\includegraphics[width=5.in]{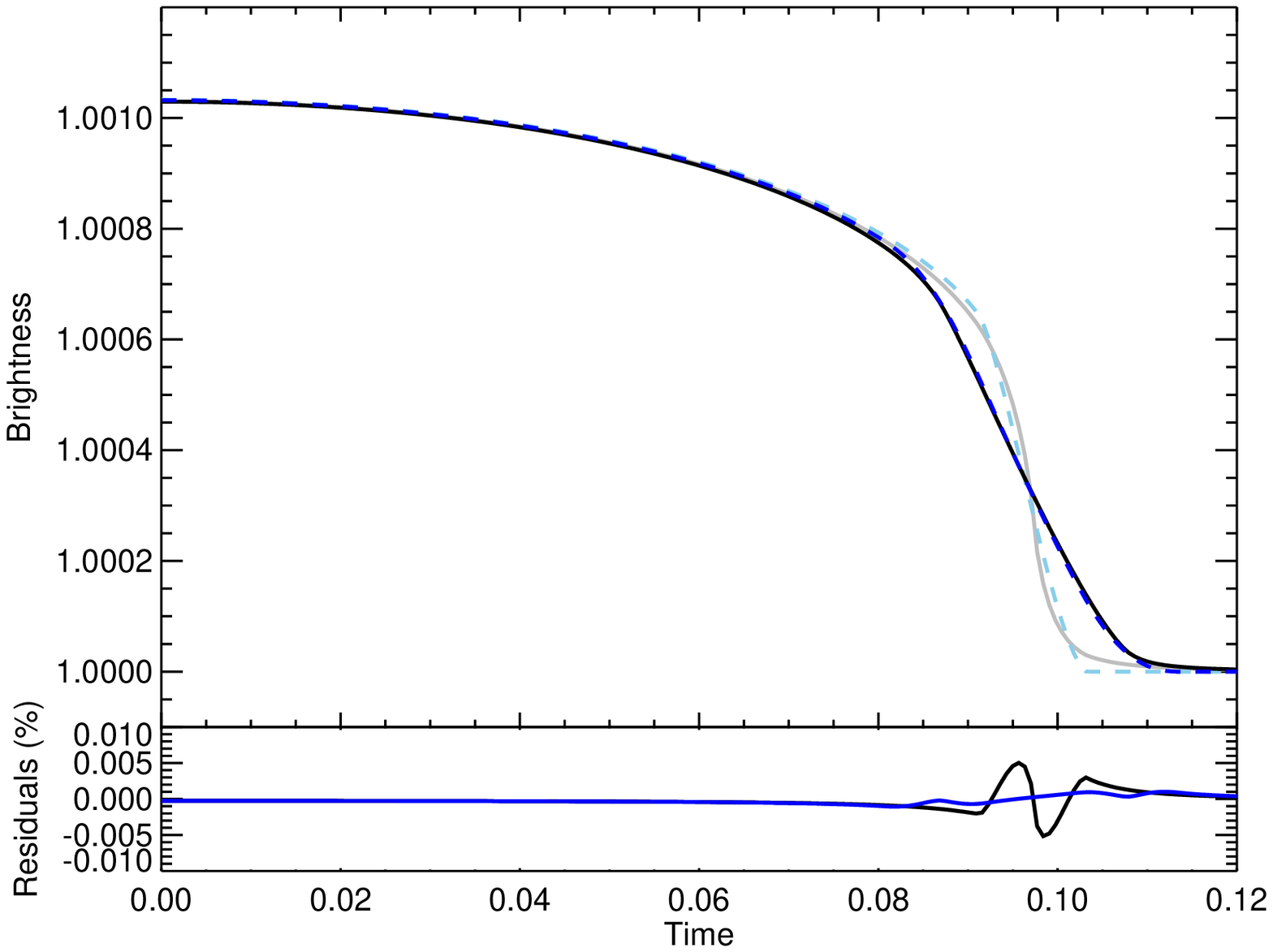}
 \caption{Comparison of the exact calculation of the microlensing pulse
\protect\cite{Agol2002} with the inverted transit approximation \protect\cite{Agol2003}
for the best-fit parameters of KOI-3278 (the pulse is symmetric,
so we only plot the second half). Top panel: exact calculation (light
grey, solid); inverted transit approximation (light blue, dashed);
exact calculation convolved with Kepler long cadence (black, solid);
inverted transit approximation convolved with Kepler long cadence
(dark blue, dashed). Bottom panel:  difference between the exact
microlensing calculation and inverted transit approximation, without
(black) and with (blue) convolution with the Kepler long cadence.
\label{fig:transit_approximation}}

\end{figure*}

We computed a light curve model for KOI-3278 (Kepler Input Catalog [KIC] number 3342467) using transiting planet modeling
software developed by one of us\cite{Mandel2002}, but with the sign of the 
flux changes switched for the microlensing pulses.  This inverted-transit approximation 
is justified because a lensing light curve shape is well approximated by 
that of a transit light curve (with small deviations at 
ingress and egress) when the Einstein radius is much smaller than the 
size of the lensed source, as is the case in this system; the only difference is a
transit's loss of light becomes a corresponding addition for a lensing event with 
the pulse height governed
 by two ratios: the Einstein radius and the lensing star's radius to the radius of the lensed source \cite{Heyrovsky1997,Agol2003}. 
In this case the ingress and egress deviations ($\lesssim 1 \times 10^{-5}$, Figure \ref{fig:transit_approximation}) are undetectable at the level of precision of the Kepler data
due to the 29.3 minute Kepler cadence.  Consequently, we utilized this
inverted transit approximation due to its much faster computation using analytic expression in terms of elliptic integrals\cite{Mandel2002}.

To an excellent approximation then, the pulse model is described by
\begin{equation} \label{pulse_model}
F(t) = F_1(t) \left(1+ \frac{2 R_E^2 - R_2^2}{R_1^2} \cdot \frac{I_1(t)}{\langle I_1\rangle}\right) + F_2
\end{equation}
where $F(t)$ is the flux from the binary system, $F_1$ is the
uneclipsed G dwarf flux, $F_2$ is the flux from the white dwarf
(assumed to be constant), $R_E$ is the Einstein radius of
the white dwarf, $I_1(t)$ is the intensity of the G dwarf 
at the location behind the center of the white dwarf, and
$\langle I_1\rangle$ is the disk-averaged intensity of the
G dwarf (this formula applies between ingress and egress).

The specific intensity of the G dwarf in the Kepler bandpass we modeled 
with a quadratic limb-darkening law;  initial fits confirmed that the signal to noise of 
the pulses was not sufficient to fit for these coefficients independently, so 
we instead adopted them from a tabulation for the Kepler bandpass as a function 
of the effective temperature, metallicity, and surface gravity of the stellar atmosphere\cite{Sing2010}.  
We fitted the tabulated limb-darkening coefficients as a function of the atmospheric
parameters, obtaining:

\begin{eqnarray}
u_1 &=& 0.4466-0.196\left(\frac{T_{eff,1}}{10^3}-5.5\right)+0.00692\log_{10}\left(\frac{g_1}{10^{4.5} }\right)+0.0865[Fe/H]_1 \cr
u_2 &=& 0.2278-0.128\left(\frac{T_{eff,1}}{10^3}-5.5\right)-0.00458\log_{10}\left(\frac{g_1}{10^{4.5} }\right)-0.0506[Fe/H]_1,
\end{eqnarray}
where $u_1, u_2$ are the linear and quadratic limb-darkening coefficients, $T_{eff,1}$ is
the effective temperature of the G dwarf in Kelvin, $g_1$ is the surface gravity
of the G dwarf in cm sec$^{-2}$, and $[Fe/H]_1$ is the abundance ratio of iron
to hydrogen, relative to the Sun, in units of dex (log base 10).  This fit is
valid in the range $5000 < T_{eff,1} < 6000$ K, $4 < log_{10}(g_1) < 5$, and
$-0.5 < [Fe/H]_1 < 0.5$, and is accurate to 0.005.  These coefficients
were used in conjunction with equation \ref{pulse_model} to compute
the light curve of the microlensing pulses, while the occultations were
computed assuming a uniform flux for the white dwarf\cite{Mandel2002}.

Since the G dwarf is spotted and undergoes quasi-periodic
fluctuations as the spots rotate in and out of view, we modeled
the $F_1(t)$ near each pulse and occultation as a quadratic function
of time and subsequently marginalized over these polynomial coefficients.
To speed up the modeling, we carried out a linearized fit for the 
polynomial coefficients of $F_1(t_i) = \sum_{j=0}^n a_j (t_i-t_j)^j$ around each event,
with $n=2$. We first computed the light curve model $F(t)$ assuming $F_1(t)=1$ 
and $F_2/F_1$ is a constant; we then divided this model into the light 
curve and solved the linear least-squares problem for the $a_j$ that minimized $\chi^2$, 
thus marginalizing over $a_j$.  This procedure ignored the 
slight variation in the ratio $F_2/F_1(t)$, but since $F_2/F_1 \approx 
10^{-3}$ and the variation in $F_1$ is a few percent at most, this error 
is of order $10^{-5}$, which is substantially smaller than the 
observational uncertainties. In computing the model, we sub-sampled each data point by a factor
of ten to properly resolve the ingress and egress of the microlensing
pulses and occultations.

We neglected photometric Doppler shift\cite{Loeb2003},
ellipsoidal variability due to tidal distortions, and reflected light from the companion star,
which are only significant for binaries with short periods\cite{Zucker2007}, and
are swamped by the stronger stellar variability in this system.
Ellipsoidal brightening can be caused by transient tidal distortion near periastron passage 
for highly-eccentric binaries known informally as ``heartbeat stars''\cite{Welsh2011,
Thompson2012};  however, these brightenings are typically asymmetric and/or show nearby 
dips, i.e.\ they do not show the inverted-U shape of the pulse seen in KOI-3278. In addition, they
typically occur near eclipse/occultation (if eclipsing) since the probability of 
eclipse is highest near periastron, while in KOI-3278 the brightening occurs opposite 
in phase to the occultation.

\subsection{Orbital model}

We modeled the orbit of the stars as a Kepler ellipse.  We used the
sky plane as the reference plane, and we refer the orbital elements
to the G dwarf, thereby defining the longitude of periastron $\omega$ as the angle 
from the point the G dwarf crosses the sky plane going away from
the observer to the G dwarf's periastron.  The 
separation between the stars, projected onto the sky, is given by:
\begin{equation}
\label{skysep}
r_{sky} = \frac{a (1-e^2)}{1+e\cos{f}} \sqrt{1-\sin^2{i}\sin^2{\left(\omega+f\right)}},
\end{equation}
where $a$ is the semi-major axis, $e$ is the orbital eccentricity,
$i$ is the orbital inclination ($i=90^\circ$ for edge-on orbit;
$i=0^\circ$ for a face-on orbit),
and $f$ is the true anomaly.  Note that since we assume a Keplerian orbit with
two bodies, the orbital elements of the two stars are the
same, with the exception of the longitude of periastra which
differ by 180 degrees. 

The microlensing pulse occurs when the white dwarf passes in front
of the G dwarf;  at this point the G dwarf has a true anomaly
which equals $f_0 = \pi/2 - \omega$.  Instead of the time of periastron
as a reference time, we use the midtime of the first pulse, $t_0$, as the 
reference time of the orbit, which is related to the time of periastron,
$\tau$, by:
\begin{equation}
\tau = t_0 + \sqrt{1-e^2}\frac{P}{2\pi}  \left[\frac{e\sin{f_0}}{(1+e\cos{f_0})} - 
2\sqrt{1-e^2} \arctan{\left(\frac{\sqrt{1-e^2}\tan{(f_0/2)}}{(1+e)}\right)}\right],
\end{equation}
where $P$ is the orbital period.  To a good approximation,
valid for small eccentricity, the time of occultation is given by $\delta t_{occ} \equiv 
t_{occ}-t_0 - \frac{P}{2} = \frac{2 P}{\pi}e\cos{\omega}$.

\section{Photometric analysis}

We carried out two independent analyses of the data:  1) separate fits to
the microlensing pulse train and to the occultations; 2) joint fits to the
pulse train and occultations using the orbital model above.  Each fit made slightly
different assumptions and used separate software; a comparison of these
two analyses for consistencey increased our confidence in each analysis
and in our inference of the parameters of the system.

\subsection{Separate fits}

The first set of fits used the Transit Analysis Package (TAP) \cite{Gazak2012}.
The pulse and occultation light curves were each computed assuming
a constant velocity and straight trajectory during each pulse/occulation
event;  this is a good approximation due to the large orbital radius and
nearly edge-on configuration.  The pulses/occulations were each fit with
five physical parameters: initial time of pulse/occultation, $t_0$; period, $P$;
impact parameter, $b$; transit duration, $T$; and the radius ratio,
$p$.  The impact parameter, $b$, is the sky-projected separation of the centers 
of the two stars at mid pulse/occultation, normalized 
to the radius of the G dwarf.  The transit duration is defined to be: 
$T = 2R_1\sqrt{1-b^2}/v$, where $v$ is the sky-velocity at 
mid-pulse/occultation.  The radius ratio, $p$, is a parameter
used in transit fitting, which is used to parameterize the
limb-darkened microlensing pulse or occultation.  In the case
of the pulse, $p = -\sqrt{\frac{2 R_E^2 - R_2^2}{R_1^2}}$,
while in the case of the occultation, $p = \sqrt{F_2/(F_1+F_2)}$.
Negative values of $p$ are converted into a flux brightening
rather than dimming, as is appropriate for the microlensing pulses.

In the TAP analysis, the limb-darkening of the G star was described
by a quadratic limb-darkening law with parameters $u_1=0.4451$ (linear)
and $u_2=0.2297$ (quadratic) which were taken from a stellar atmosphere 
model with $T_{eff}=5500$ K, $log(g)=4.5$, and $[Fe/H]=0.0$\cite{Sing2010}.  
The white dwarf was assumed to have no limb-darkening;  this is a sufficient 
approximation since only the ingress/egress of the occultation is sensitive
to the white dwarf limb-darkening, while this portion of the light
curve has very low signal-to-noise due to the long Kepler cadence. In 
addition to the quadratic variation of the G dwarf, a correlated-noise 
model assuming $1/f$ noise in addition to a white-noise component was
solved for along with the model parameters\cite{Carter2009}.  The
red and white noise amplitudes were allowed to vary separately for
each pulse and occultation.

Table \ref{tab:lightcurve} shows the results of the TAP fits;  we
fit the posterior of these parameters with a Gaussian, and report
the mean and standard deviation of the Gaussian fits (each set
of parameters was weakly correlated).  

\begin{table}
\begin{center}
\caption{Light curve parameters \label{tab:lightcurve}}
\begin{tabular}{lc}
\hline \hline
\multicolumn{2}{c}{\it Pulses:}\\
\hline
Period (d)& $ 88.18025\pm  0.00049$ \cr
Duration (d)& $ 0.1955\pm 0.0043$ \cr
Pulse height $D$& $ 0.00102\pm 0.00005$ \cr
$t_0 + 7 \times P$ (JD-2,454,900)& $702.68181\pm  0.00196$ \cr
\hline
\multicolumn{2}{c}{\it Occultations:} \\
\hline
Period (d)& $ 88.18091\pm  0.00028$ \cr
Duration (d)& $ 0.1914\pm 0.0027$ \cr
Occultation depth & $ 0.00112\pm 0.00004$ \cr
$t_0 + 7.5 \times P $ (JD - 2,454,900) & $703.50886\pm  0.00124$ \cr
\hline
\hline
\end{tabular}
\end{center}
\end{table}

The ephemerides of the microlensing pulse train and the series
of occultations were fit separately; we found that their periods
were nearly identical, $P_{pulse}=88.18025\pm 0.00049$ d and
$P_{occ} = 88.18091\pm 0.00028$ d, for a difference of
$P_{pulse}-P_{occ} = -0.9 \pm 0.8$ min.  This indicates that
both the microlensing pulse train and occultations can be described by
a single, Keplerian orbital model, which justifies the joint fit
in the next section.  The occultation occurs $\Delta t_{occ} =
t_{0,occ}-t_{0,pulse}-P/2 = 0.827 \pm 0.002$ days later than half of
an orbital period after the pulse.  This translates into $e\cos{\omega} 
= 0.01473\pm 0.00004$.

We also found that the impact parameters of the pulse train
and occultations were poorly constrained;  only near-grazing
configurations could be excluded.  This is due, once again,
to the long-cadence data which place no constraint on the
ingress/egress duration, approximately $6-9$ min, which is shorter than 
the 29.3-minute Kepler cadence.  For the occultation, the TAP model 
was incorrect at ingress/egress in that it fixed the depth to equal the square root of 
the radius ratio; consequently we do not trust the impact parameter 
constraint on the occultation.  For the pulse, the impact parameter 
likelihood showed a decline above an impact parameter of $b=0.65$, 
which we fit with a linear decline down to $b=1.06$.

We found that the durations of the pulses and the occultations
were identical to within the errors, $T_{pulse} = 0.196\pm 0.004$ d
and $T_{occ} = 0.191\pm 0.003$ d, for a difference of $T_{pulse}-
T_{occ} = 6\pm 7$ min and a ratio of $T_{pulse}/T_{occ} = 
1.02\pm0.03$.  A ratio near unity also indicates that both events can be described 
by a single Keplerian orbital model with (likely) small $x= 
e\sin{\omega}$. In theory the ratio of the durations can be
used to constrain $e\sin\omega$;
to lowest order in $x = e\sin{\omega}$:
$\frac{T_{pulse}}{T_{occ}} = 1 + a x$, with $a = 2\frac{2 b_0^2 (1-y^2) 
-1}{1-b_0^2(1-y^2)^2}$, $y=e\cos{\omega}$ and
$b_0 = a\cos{i}/R_1$ (the impact parameter if $e=0$).
In the limit $b_0=0$, $a=-2$, which yields $e\sin{\omega} = -0.0035\pm
-0.016$.  However, at larger impact parameter the sign of
$a$ switches, and $a$ goes to zero for $b_0=1/\sqrt{2}$; near this 
value the $x^2$ term which we 
have dropped in $T_{pulse}/T_{occ}$ becomes important in constraining the
value of $e\sin{\omega}$. The best-fit impact parameter from
the joint model gives $b_0=0.706\pm0.022$; this translates to
$a=-0.1\pm$0.5, which spans zero, so $e\sin{\omega}$ has a larger
uncertainty than this linear expansion estimate;  it is properly constrained
by the full Markov chain solution.

To translate the separate constraints on the shape of the
light curve into constraints on the system parameters, we next
carried out an MCMC analysis using analytic formulas to
describe the light curve parameters in terms of the masses,
radii, and orbital parameters of the G dwarf and white
dwarf\cite{Winn2011}.  We found a strong correlation between
$T$, $p$ and $b$ for the microlensing pulse fit, so
we reparameterized $p$ as $D(b)=p^2 \frac{1-u_1\mu-u_2 \mu^2}
{1-u_1/3-u_2/6}$ where $\mu = 1-\sqrt{1-b^2}$.  This parameter
approximates the maximum height of the microlensing pulse, at
its center, and is nearly uncorrelated with
$b$ and $T$.  Although the TAP light curve fits held $u_1$ and 
$u_2$ fixed, we allowed these to vary during this second step of 
the analysis computing $D(b)$.  We also transformed the zero-points
of the ephmerides to points near the middle of the series
of pulses/occulations so that they were uncorrelated with
the orbital periods.  The results of these fits were used
for rapid experimentation with various assumptions in our
analysis and comparison with the joint fits described next;
however, the joint fits have the advantage of self-consistently
fitting all of the data simultaneously, so we use the joint
fits for our final parameter constraints.

\subsection{Joint fits}
\label{jointfits} 

The second model jointly simulated the pulses and occultations and compared to 
the observed Kepler fluxes near the events. Using the masses of the two bodies 
and their orbital elements as inputs ($M_1, M_2, P, t_0, i, e, \omega $), we 
calculated the two stars' projected sky separation (Equation \ref{skysep}) at all 
times of interest (i.e. Kepler cadences surrounding the pulses and occultations, 
subsampled by a factor of 10).  We combined this sky separation with the radii of 
the two stars, their flux ratio, and the G-dwarf's limb darkening coefficients 
($R_1, R_2, \frac{F_2}{F_1}, u_1, u_2$) to predict the observed flux at each 
cadence using the method of \S\ref{lcmodel}; we then calculated the $\chi^2$ of the
model with these 12 parameters.

To further constrain the system parameters and break degeneracies between them, 
we utilized stellar evolution models and added photometric constraints from other 
surveys to our $\chi^2$ calculation, which required adding additional input parameters (\S\ref{sec:stellar_parameters}). The Kepler light curve 
does not have high enough signal to noise to constrain certain inputs 
(e.g. $u_1, u_2, R_2$), so we reparametrized them as a function of more accessible inputs 
(\S\ref{reparam}).  We then used an MCMC analysis on our final set of 14 system parameters 
to determine the stellar and orbital properties and their posterior distributions.

\subsubsection{Photometric constraints on stellar parameters}
\label{sec:stellar_parameters}

\begin{figure*}
\center
\includegraphics[width=5.in]{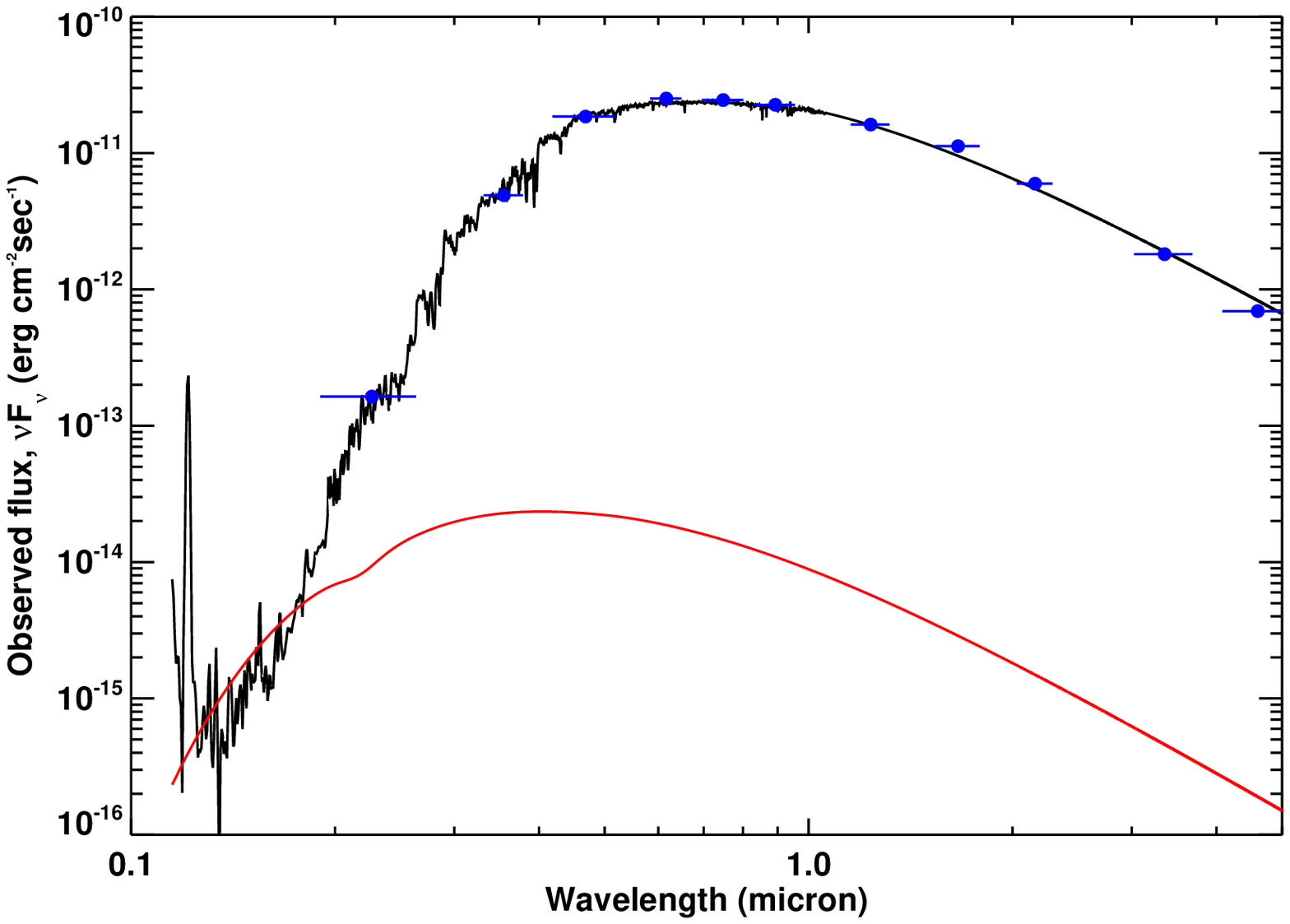}
 \caption{Spectral energy distribution of KOI-3278, computed from GALEX, SDSS, 2MASS,
and WISE photometry (blue);  the two shortest wavelength bands were not used in our fitting.
Overplotted in black for illustration is a Pickles composite spectrum
\protect\cite{Pickles1998} for a G5V star ($T_{eff} = 5584$ K), with extinction applied, 
as well as a blackbody at 9950 K with the best-fit size-ratio estimated for the white 
dwarf (red). \label{fig:sed}
}
\end{figure*}

Because we do not have spectroscopic data for this system, we relied
on multi-wavelength photometry and stellar evolution models for 
computing the stellar properties.  A determination 
of the stellar characteristics based on multi-color photometry using the Dartmouth 
stellar evolution models \cite{Dotter2008} has already been carried out by the Kepler team
\cite{Pinsonneault2012,Huber2013}.  However, these analyses have several
drawbacks:  they assumed priors on the temperture, metallicity, and mass 
based upon the properties of stars in the Solar neighborhood; they assumed a 
simplistic model for the extinction/reddening correction; and the covariances between
the resulting parameters were not reported.  Instead, we carried out our own fits, using
a simultaneous $\chi^2$ minimization of the Kepler photometric light curve and
 multi-band photometry (Figure \ref{fig:sed}) from SDSS $g,r,i,z$ \cite{Abazajian2009}, 2MASS
$J,H,K_s$ \cite{Cutri2003}, and WISE $W1, W2$ \cite{Cutri2012} in order to provide 
joint constraints on the properties of both stars and their orbital properties.

\begin{figure*}
\center
\includegraphics[width=5.in]{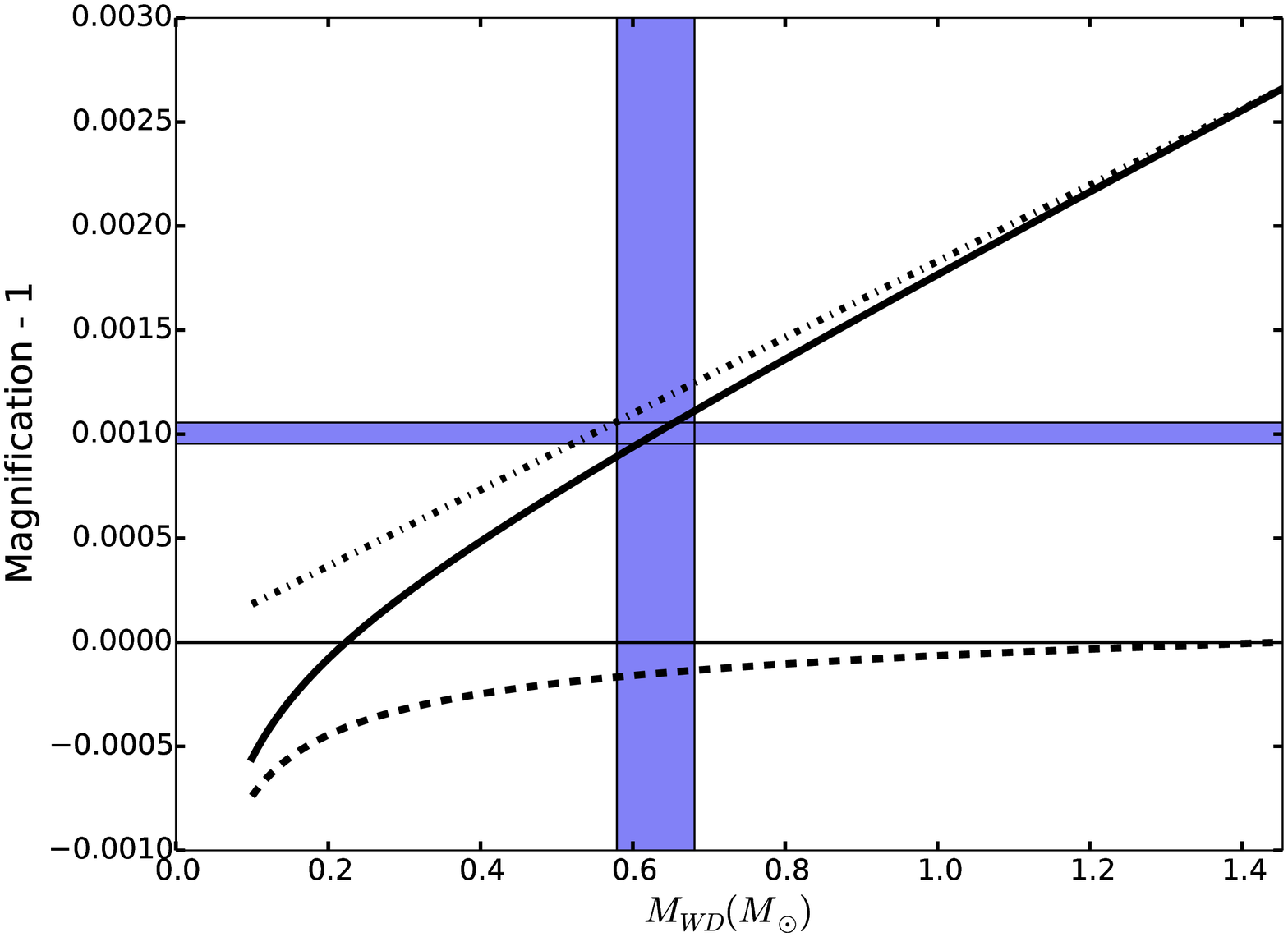}
 \caption{The ``magnification'' (minus one) versus mass of the white dwarf,
neglecting limb-darkening of the G dwarf.
The dashed-dotted curve shows the magnification versus white
dwarf mass (i.e.~assuming the white dwarf is transparent);  the 
dashed curve shows the eclipse depth of the white dwarf
(i.e.~traditional eclipse ignoring lensing; remember more massive white dwarfs
have smaller radii); and the black curve
shows the predicted pulse height balancing the two effects.  
The blue regions show the
$1\sigma$ uncertainties on the measured values. \label{fig:pulse_height}}

\end{figure*}

This method has the advantage of self-consistently accounting for all of the stellar 
properties simultaneously, as well as taking into account the covariances between
stellar evolution model parameters.  For example, the pulse/occultation
duration is a function of the density of the G dwarf and the
ratio of the total binary mass to the G dwarf mass; the G dwarf density strongly 
correlates with the effective temperature of the star in this temperature range.  
Also, the height of the microlensing pulse primarily constrains the
mass of the white dwarf star, given the parameters for the G dwarf
and orbit (Figure \ref{fig:pulse_height}).
By fitting the multi-color photometry and 
light curve simultaneously we obtained a self-consistent fit to all of these constraints
on the G dwarf, white dwarf, and orbital elements.

The photometric fit in multiple bands required correction for reddening.
The total extinction, 
$A_{\lambda,max}$, we estimated from reddening maps of the galaxy\cite{Schlafly2011}.
We assumed $E(B-V)_{max}$ has a fractional uncertainty of 3.5\% based on the scatter of
nearby pixel elements in the extinction map, and fixed $R_V = 3.1$. 
We then corrected for the finite extent of the dust layer by adding a free scale height parameter with
a prior of $h_{dust} = 119\pm 15$ pc\cite{Jones2011}.  The correction for the extinction column 
becomes: $A_\lambda = A_{\lambda,max} (1-\exp{\left[-D \sin{10.29^\circ}/h_{dust}\right]})$, 
where $D$ is the distance to the binary in parsecs -- another free parameter added to the model. 
Finally, we added a systematic uncertainty in the absolute calibration of 
the photometry, $\sigma_{sys}$, which we added in quadrature to the reported photometric 
errors of the measured magnitudes.  We let $\sigma_{sys}$ vary as a free parameter,
and placed a prior on its value of $\prod_i^N (\sigma_i^2+\sigma_{sys}^2)^{-N/2}$ 
where N=9 is the number of photometric bands;
this has the effect of giving a reduced $\chi^2$ of order unity
for the photometric fit.  The median value of $\sigma_{sys}$ was 2.5\% in our fits.

Fitting the observed broadband magnitudes in addition to the 
Kepler light curve therefore required adding $D$, $\sigma_{sys}$, $h_{dust}$, and
$E(B-V)_{max}$ as free parameters

For modeling the SED of the G dwarf, we used the Padova PARSEC isochrones \cite{Bressan2012}, 
with scaled solar alpha abundances ([$\alpha$/Fe] = 0).  We used this publicly available 
grid of stellar models computed for ages from $0.004 < t_1 < 12.59$ Gyr (spaced by
0.05 dex), metallicities from $-1.8 <[Fe/H]_1< 0.7$ (spaced by 0.1 dex), and masses from 
$0.1 < M_1 < 11.75 M_\odot$ (with spacings depending on age and metallicity, adaptively
chosen by the isochrone model).  By utilizing $M_1, [Fe/H]_1, t_1$ to parameterize our Markov 
chain fits, we place a uniform prior on these parameters.  We carried out linear interpolations 
of these parameters in the grid of stellar models to compute the radius ($R_1$), effective 
temperature, $T_{eff,1}$, $log(g_1)$, and absolute magnitudes of the 
G dwarf for comparison to the multi-band photometric data, and for computation of the 
light curve model (the age interpolation we carried out linearly in $\log_{10}(t_1)$, although 
we used $t_1$ as the Markov chain parameter in order to avoid favoring small ages).

We checked the robustness of our results by redoing the fits with
the Dartmouth isochrones\cite{Dotter2008}.  Unfortunately the Dartmouth
isochrones have coarse sampling in metallicity below Solar metallicity
(0.5 dex), so our interpolation fared poorly for sub-solar metallicity.  
Instead we re-ran our fits with only positive metallicity (using the
separate fits described above), and we found 
that we obtained statistically identical results for both the Padova and
Dartmouth isochrones with the constraint of super-solar metallicity.  We 
conclude that our results are robust to the choice of isochrone;  this is not
surprising as the G dwarf star is near solar mass, where stellar evolution
models are robustly constrained by comparison with our Sun. Note that the Dartmouth
and Padova isochrones assume slightly different metallicities for the Sun 
($Z=0.019$ and $Z=0.0147$, respectively), which we accounted for in our comparison.

\subsubsection{Reparametrization}
\label{reparam}
 
Some inputs to the light curve modeling are highly correlated (e.g. $e, \omega$), while 
others are poorly constrained by the data due to the long Kepler cadence or low signal 
to noise (e.g. $u_1, u_2$). We therefore reparametrized our model inputs into more insightful 
and independent parameters.
 
As discussed in \S\ref{lcmodel}, the limb darkening coefficients of the G-dwarf cannot be 
constrained by the data; $u_1, u_2$ are thus transformed into dependent functions of the G-dwarf, 
which are in turn determined by the isochrones and input parameters $M_1, [Fe/H]_1, t_1$. The isochrones 
similarly determine $R_1$, and it is no longer treated as a free parameter.

We reparameterized the inclination angle in
terms of the impact parameter of the white dwarf during the microlensing
pulse if the orbit were circular, $b_0 \equiv (a/R_1) \cos{i}$. 

We transformed the eccentricity and longitude of periastron
to $e\sin{\omega}$ and $e\cos{\omega}$ since these are better
characterized than $e$ or $\omega$ alone; this change
requires placing a prior of $1/e$\cite{Eastman2013}.

Since we could not constrain the white dwarf radius from these data, we
assumed a mass-radius relation for the white dwarf given by 
\begin{equation}
R_2(M_2) = 0.0108 R_\odot \sqrt{\left(\frac{M_2}{M_{ch}}\right)^{-2/3}-
\left(\frac{M_2}{M_{ch}}\right)^{2/3}},
\end{equation}
where $M_{ch}=1.454 M_\odot$ is the Chandrasekhar mass.

In our final fits, we constrained the age of the system to be the sum of the
main-sequence lifetime of the white dwarf progenitor and white 
dwarf cooling time, which amounts to exchanging $F_2/F_1$ for
the mass of the white dwarf progenitor, $M_{2,init}$ (see \S\ref{sec:MSage}).
Ultimately, the final set of parameters we fit for were:
\begin{equation}
\left\{ P,t_0,e\sin{\omega},e\cos{\omega},b_0,M_2,M_{2,init},M_1,t_1,[Fe/H]_1, \sigma_{sys},D,h_{dust},E(B-V)_{max} \right\}
\end{equation}
for a total of fourteen free parameters.  

\subsection{Results}

Our initial joint fits gave a reduced chi-square slightly larger than unity,
so we increased the Kepler photometric error bars by a factor of 1.13 
in the joint fits to produce a reduced 
chi-square of unity in our fit to the Kepler time-series photometry.

We ran a Markov Chain Monte Carlo simulation to constrain
the fourteen model parameters using an ensemble sampler with
affine-invariance\cite{Goodman2010,ForemanMackey2013}. We
used a population of 50 chains and ran for a total of
100,000 generations, with maximum Gelman-Rubin statistic of 1.06.

\begin{figure*}
\center
\includegraphics[width=7.5in]{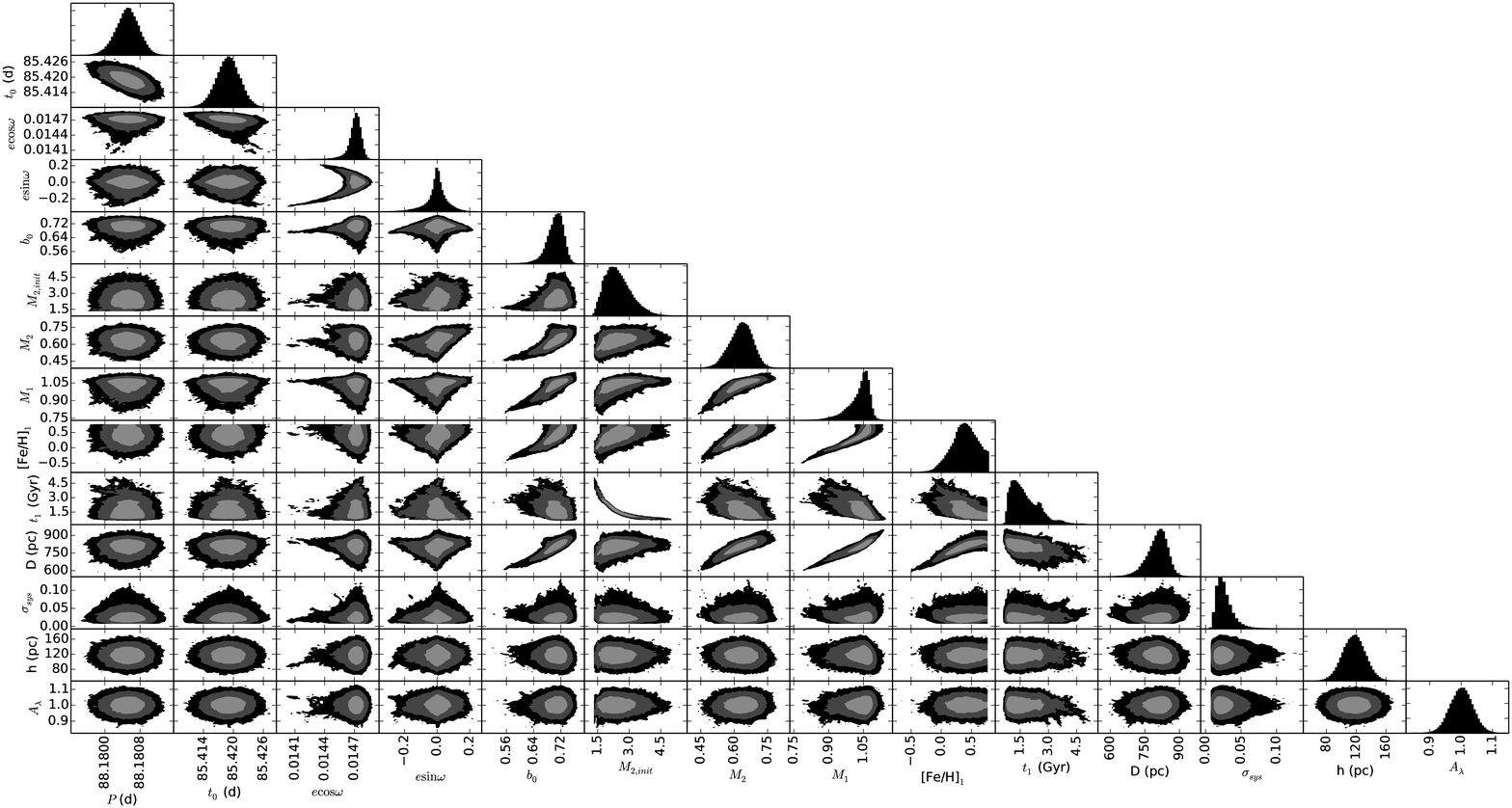}
 \caption{Contour plots showing the $1\sigma$, $2\sigma$, and $3\sigma$
constraints on pairs of parameters. \label{fig:triangle}}
\end{figure*}

Table 1 lists the resulting parameters derived from our simulations. Some 
parameters have extremely strong correlations; in particular, the measurement 
of the mass of the white dwarf is limited by our uncertainty in the model of 
the G dwarf star.  Figure \ref{fig:triangle} shows the correlations between
various model parameters.

\section{Age Constraints}

During our initial fits, we found a correlation between
the age of the G dwarf and the mass of the white dwarf.  This
can be understood as follows:  the multi-band photometry
constrains the effective temperature of the G dwarf.  As
stars evolve, they expand in size, but a larger radius for
the G dwarf requires a larger mass of the white dwarf to reach 
the same microlensing pulse magnification (which scales
as $M_2/R_1^2$).  In addition, the larger radius of the
G dwarf causes a longer transit duration;  to fit the
observed duration requires a higher impact parameter where the star is dimmer, 
which also works to increase
the white dwarf mass needed to reach the same pulse height.

This leads to a problem with the age of the binary.
To produce the observed flux ratio between the stars
(derived from the occultation depth) requires 
a recently formed white dwarf.
Yet older G dwarfs with larger radii require higher white dwarf
masses to match the pulse heights; more massive
white dwarfs are created by higher mass stars which have
shorter main-sequence lifetimes. Hence older G dwarfs
require both a short main-sequence lifetime for the white dwarf
progenitor as well as a young white dwarf, producing a binary with
contradictory stellar total ages.

We thus eliminated the high-mass WD and old G-dwarf solutions
by requiring our binary system to be coeval. We constrained the minimum age with
the spin period of the G dwarf (\S\ref{sec:rotation}), and constrained the 
maximum age with an initial-final mass 
relation for the white dwarf, which determines the nuclear-burning 
lifetime of the white dwarf progenitor (\S\ref{sec:MSage}).

\subsection{Period of rotation of G dwarf and spin-down age}
\label{sec:rotation}

\begin{figure*}
\center
\includegraphics[width=5.in]{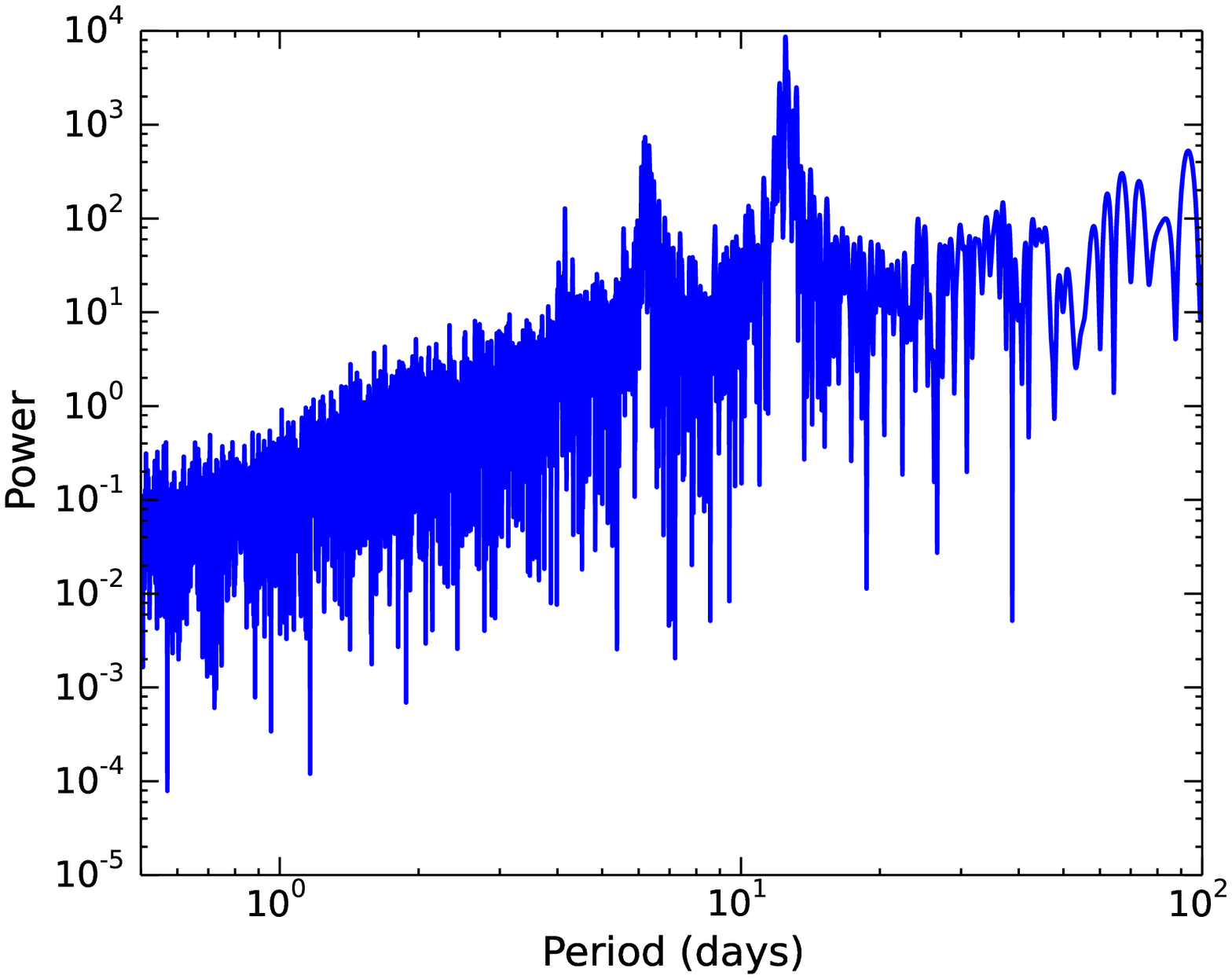}
 \caption{Frequency-power spectrum of KOI-3278, showing a strong peak
at 12.5 days which we infer to be the rotational period of the G dwarf. \label{fig:powspec}}
\end{figure*}

The light curve of KOI-3278 looks like a typical spotted star with
star spots repeating every $\approx$12 days;  a power spectrum peaks
strongly at 12.5 days (Figure \ref{fig:powspec}).  Using only Q3 data, the period of 
rotation was measured by Reinhold et al.\cite{Reinhold2013}, in 
which they report a best-fit period of 12.36$\pm$0.05 days, consistent 
with our results from all 17 quarters.

The rotation period can be used to estimate the age of the G dwarf.
We estimated the age of the star based on
the observed spin-down of stars as they age; so-called ``gyrochronology.''
We used the calibrations of this relation by \cite{Mamajek2008} 
to estimate the age of this star, which we found to be $t_{spin,1} = 
0.89 \pm 0.15$ Gyr. We used this constraint only as the minimum age of the system, however,
to allow for the possibility that the G-dwarf was spun up via mass 
transfer during the white dwarf's formation.

\subsection{Breaking the $M_2$, Age Degeneracy}
\label{sec:MSage}
To eliminate the models with contradictory ages, we placed a constraint on the age of the G dwarf by adding
together the cooling age of the white dwarf and the nuclear-burning
lifetime of its progenitor.  However, the progenitor mass has some
uncertainty due to uncertainties in the initial-final
mass relation of white dwarfs, which is compounded by the
fact that the common-envelope evolution of this system would
have modified the core mass of the white dwarf progenitor.
After surveying the literature on the initial-final mass
relation of white dwarfs\cite{Weidemann2000,Kalirai2008,
DeMarco2011,Catalan2008,Zhao2012}, and running a suite of binary
stellar evolution models (described below), we found that
most data and models lay within 10\% of the final mass
given by the initial-final mass relation found by 
Kalirai\cite{Kalirai2008}.  Consequently, we allowed both the 
initial mass of the white dwarf, $M_{2,init}$, and the final mass, 
$M_2$, to vary, and placed a Gaussian prior on $M_2$ to lie within
10\% of the Kalirai relation, which amounts to adding
to the $\chi^2$: $(M_2 - 0.109 M_{2,init} - 0.394)^2/(0.1 M_2)^2$.

We computed the nuclear-burning lifetime of the white
dwarf progenitor, $t_2$, from the Padova models, and then set the cooling 
time of the white dwarf equal to $t_{cool} = t_1 - t_2$.  The
cooling time and mass of the white dwarf was then used to
compute its Kepler magnitude (as described below), which
was then used to fit the depth of the occultation.  This procedure
has the effect of requiring both stars to have the same age,
but allowing for some uncertainty in the initial mass of the
white dwarf progenitor.  In doing so, we exchanged $F_2/F_1$
for $M_{2,init}$ as a free parameter in the model.  This
procedure eliminated the unphysical cases of large white
dwarf masses in old systems.

\subsubsection{WD Cooling}

We can derive the
absolute magnitude of the white dwarf star in the Kepler
band from the flux lost as the white dwarf completely disappears behind the G-dwarf during occultation;  we can then use this to constrain the age and
luminosity of the white dwarf based on white dwarf
cooling models.  We use the cooling models computed by
Bergeron and collaborators\cite{Holberg2006,Kowalski2006,
Tremblay2011,Bergeron2011}, made available on
their web site\footnote{{\tt http://www.astro.umontreal.ca/$\sim$bergeron/CoolingModels/}}.  
We performed a linear interpolation in the mass and log cooling age of
the white dwarf to obtain the absolute magnitudes, luminosity, 
and effective temperature of the white dwarf.  The absolute 
magnitude of the white dwarf in the Kepler band was computed by
transforming the absolute magnitudes in the SDSS $g, r$ and
$i$ bands: $K_p=0.25g+0.75r$ for $(g-r) \le 0.3$ and
$K_p = 0.3g+0.7i$ for $(g-r) > 0.3$\cite{Brown2011}.

We found a white dwarf age of $t_{cool} = 663\pm60$ Myr; equivalent 
ages were found for both Helium and Hydrogen atmosphere models.  The effective
temperature of the white dwarf is $T_{eff,2} = 9960\pm730$ K and its
luminosity is $L_{WD} = (1.2\pm0.23)\times
10^{-3} L_\odot$; hence the small depth of the occultation in
the Kepler band.

\section{Blends}

It is possible that the flux from another star (or stars) can be blended 
with the flux of the binary star, thus affecting our fit to the
photometry and light curve.  To test this, we added to our separate-fit
model two components: 1) a bound blend star with the same metallicity and 
age as the G dwarf; 2) a blend star along the line of sight to the binary,
contained within the Kepler photometry aperture.

For the first component, we added the flux of a second star to
the multi-color photometry, drawing from the Padova isochrones
as for the G dwarf, and we also included its effect on the pulse
height and occultation depth.

For the second component, the contamination within the Kepler
aperture can be estimated by combining the location of other stars in
other photometric surveys with the Kepler point spread
function to compute the flux contamination with the target aperture.  
The Kepler pipeline carries out this analysis, and finds that the
contamination is between 4-8\%, depending on the quarter of
data that is used.  We added the contamination flux to both our models
to account for the slight reduction in the pulse height/occultation
depth due to contamination that varies with quarter.

We re-ran the Markov chain fit including the mass of the third bound
star, $M_3$, as an additional free parameter.  We found that a bound star must
be an M dwarf, $M_3 = 0.4\pm0.2 M_\odot$, to be consistent with the 
data, and would contribute only $1.4^{+3.7}_{-1.0}$\% to the Kepler band 
flux.  However, the M dwarf would contribute more significantly to the 
2MASS/WISE bands and thus skew the effective temperature of the G 
dwarf to be somewhat hotter, and thus somewhat more massive, 
$M_1=1.07^{+0.04}_{-0.05} M_\odot$.  This would imply a slightly higher white 
dwarf mass, $M_2=0.68^{+0.05}_{-0.06} M_\odot$, about $1\sigma$
different from the fit without a third star.  The slightly higher mass for
the white dwarf would produce a slightly higher mass for its progenitor,
as well as a slightly smaller cooling age, and thus a slightly
smaller overall age for the system.  Higher contrast imaging 
and/or high resolution spectroscopy may be able to place stronger 
constraints on the presence of a third bound star in the system;  
however, the current constraints are strong enough that the mass 
derived for the white dwarf is not strongly affected by the third star.

\section{Binary stellar evolution models and dynamical constraints on 
the presence a bound third star}

We carried out an initial exploration of the possible orgin of
this system using the {\sc BSE} code \cite{Tout1997,Hurley2000,Hurley2002} 
to model the evolution of this system as a function of time.   At the
current orbital period, the G dwarf should have orbited within the 
surface of the red giant progenitor of the white dwarf, ejecting
the outer envelope of the star;  this is referred to as the
``common envelope phase''\cite{Paczynski1976,Iben1993}.  We used the 
calibration of the common-envelope evolution parameters derived empirically\cite{DeMarco2011},
and carried out simulations with a range of initial masses and separations.
As an example of these simulations, we found that the final conditions of this 
system could be achieved if the initial masses were $M_{2,init} \approx 2.5 M_\odot$, 
$M_1 \approx 1 M_\odot$ and the initial period was $P_0 \approx 1295$ days, 
corresponding to an initial semi-major axis of $a_0 \approx 3.5$ AU (we used
$\alpha_{CE} = 0.3$ and $\lambda = 0.2$ in this simulation).  The
mass transferred during the Roche-lobe overflow and common envelope phase
would be $\Delta M_1 = 0.008 M_\odot$, sufficient to spin up the G dwarf.
The common envelope phase would start on the second asymptotic giant branch
of the white dwarf progenitor, at an age of $\approx 0.8$ Gyr, and result in 
a final mass of the white dwarf of $M_2 \approx 0.65 M_\odot$, consistent 
with the model constraints, and slightly less massive than the final mass
of $0.69 M_\odot$ had the star evolved as a single star.  As the common
envelope phase causes rapid merging of the two stars before ejection
of the evolved star's envelope, the final period is very sensitive to
the initial period;  the outer period thus has to be fine-tuned, and
hence this sort of binary is expected to be rare\cite{deKool1993,deKool1995,
Green2000,Schreiber2003,Willems2004,Zorotovic2010,Davis2010,Toonen2013}.

The possibility of a third body in the system is potentially constrained by 
the dynamical interactions of the 3 bodies due to the eccentric-Kozai
mechanism\cite{Ford2000,Katz2011,Naoz2013}.  The observed eccentricity of the binary is small,
$e_1 \approx 0.032$, which indicates that it was probably circularized
during the common-envelope phase, and avoided dynamical growth of
its eccentricity with a third body, post-circularization.  Since
the timescale for growth of the eccentricity depends upon the
quadrupole timescale, we estimate that the third body should
satisfy $(a_2/AU)^3/(M_3/M_\odot) > 1.7 \times 10^8$ so that
the Newtonian quadrupole timescale is less than the white
dwarf cooling timescale.  Thus, if the third body has a mass of 
$M_3 \approx 0.4 M_\odot$, the semi-major axis should be larger than 
$a_2 > 748$ AU, with an orbital period longer than 14,000 yr.  This is 
at about a separation of 1$^{\prime\prime}$ (at quadrature), and thus 
the presence of a third body could be constrained with future high-contrast 
imaging and dynamical simulations.

\begin{figure*}
\center
\includegraphics[width=5.in]{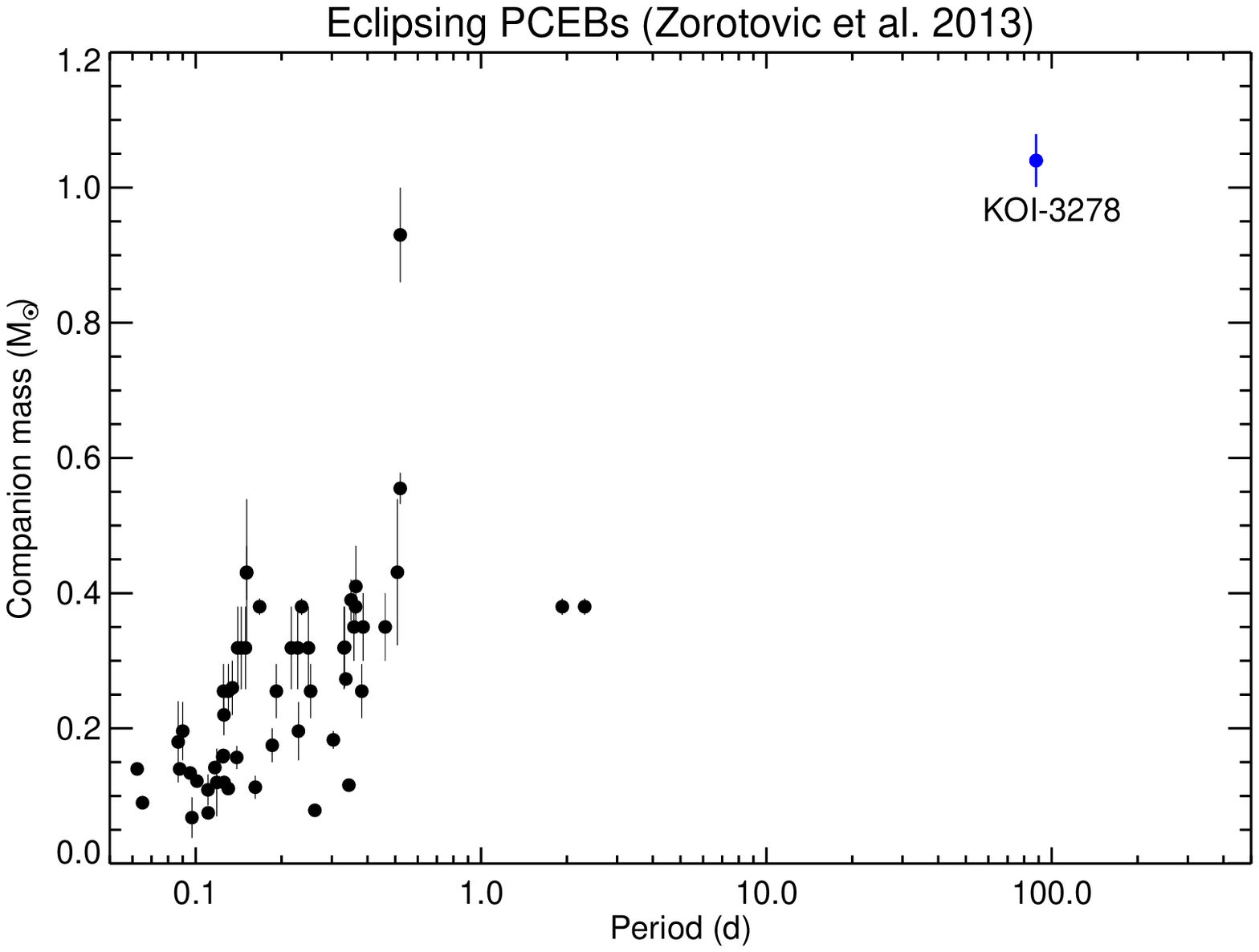}
 \caption{Mass-period distribution of known white dwarf-main
sequence post common envelope eclipsing binaries. \label{fig:period_mass_pceb}}

\end{figure*}

The value of this system can be seen when comparing with the other
white dwarf-main sequence eclipsing binaries found to 
date (Figure \ref{fig:period_mass_pceb}).  KOI-3278 has the most massive companion star, as well
as the longest period of all such systems.  The longer period binaries
are more difficult to find with ground-based surveys, and also
have a lower probability of eclipse/occultation/microlensing.
The cooler companion stars are easier to find due to their
larger color difference to their companion white dwarf stars.
KOI-3278 could only be found with continuous coverage and
with high photometric sensitivity.

\section{Predictions for future observations}

Based on our Markov chain analysis, we found that the velocity
semi-amplitude of the G dwarf should be $K_1 = 21.5 \pm 1$ km/s.  The parallax
of the system should be $\pi = 1.24^{+0.08}_{-0.05}$ milliarcseconds (mas) with a reflex motion
 of $\alpha = 0.22\pm 0.08$ mas.
The expected parallax measurement uncertainty for a $G=15$ (Gaia magnitude) 
star is 0.02-0.03 mas\cite{deBruijne2012}, so the parallax precision from Gaia
should improve upon our analysis significantly, and enable another
constraint on the mass of the white dwarf star.

We simulated the flux ratio of the white dwarf to the G
dwarf as a function of wavelenth, which we find reaches $\approx 5$\% at 0.25 micron,
increasing to 60\% at 0.15 micron (although the absolute flux
drops significantly towards shorter wavelength).  We found that
single occultation measurements in the ultraviolet could have similar 
signal-to-noise as the combined Kepler occultations, and allow 
a measurement of the temperature of the white dwarf, breaking 
the size-temperature degeneracy that required us to use a 
mass-radius relation for the white dwarf.  For instance,
we found that observations using the Hubble Space Telescope at 0.2-0.4 micron (with
the G280 grating on the Wide Field Camera III) could achieve
a S/N of $\approx 25$ with observation of a single occultation
(if it were photon-noise limited).

{\bf Acknowledgments:} We acknowledge the dedication and hard work of the Kepler team in obtaining and analyzing the data used in our analysis. This research has made use of the VizieR catalogue access tool, CDS, Strasbourg, France.   This research has made use of NASA's Astrophysics Data System.

\end{document}